\documentclass{aa}
\usepackage{epsfig}
\usepackage{graphics}
\usepackage{subfig}
\usepackage{natbib}
\usepackage{amsmath}
\usepackage{enumerate}
\usepackage{threeparttable}
\usepackage{caption}

\begin{document}

\title{AGN jets versus accretion as reionization sources}

\author{ N\'uria Torres-Alb\`a\inst{1}, Valent\'i Bosch-Ramon\inst{1}, and Kazushi Iwasawa\inst{1,2}}
  
\institute{Departament de F\'{i}sica Qu\`antica i Astrof\'{i}sica, Institut de Ci\`encies del Cosmos (ICC), Universitat de Barcelona (IEEC-UB), Mart\'{i} i Franqu\`es 1, E08028 Barcelona, Spain
\and ICREA, Pg. Llu\'is Companys 23, E-08010 Barcelona, Spain}

\offprints{N. Torres-Alb\`a \\ \email{ntorres@fqa.ub.edu}}

\titlerunning{Reionization from AGN jets}


\abstract
{Cosmic reionization put an end to the dark ages that came after the recombination era. Observations seem to favor the scenario where massive stars generating photons in low-mass galaxies were responsible for the bulk of reionization. Even though a possible contribution from accretion disks of active galactic nuclei (AGN) has been widely considered, they are currently thought to have had a minor role in reionization.}
{Our aim is to study the possibility that AGN contributed to reionization not only through their accretion disks, but also through ionizing photons coming from the AGN jets interacting with the IGM.}
{We adopt an empirically derived AGN luminosity function at $z\simeq6$, use X-ray observations to correct it for the presence of obscured sources, and estimate the density of jetted AGN. We then use analytical calculations to derive the fraction of jet energy that goes into ionizing photons. Finally, we compute the contribution of AGN jets to the H II volume filling factor at redshifts $z\simeq15-5$.}
{We show that the contribution of the AGN jet lobes to the reionization of the Universe at $z\sim6$ might have been as high as $\gtrsim 10$\% of that of star-forming galaxies, under the most favorable conditions of jetted and obscuration fraction.}
{The contribution of AGN to the reionization, while most likely not dominant, could have been higher than previously assumed, thanks to the radiation originated in the jet lobes.}
 
 
\maketitle

\section{Introduction}

Cosmic reionization represents an important stage in the evolution of the Universe, putting an end to the dark ages that came following the recombination era. Observations indicate that the intergalactic medium (IGM) was completely reionized at redshift $z\simeq 6$ \citep[e.g.][]{FanStra2006,PenVan2014,TilPap2014,McGMes2015}. However, the onset and duration of reionization remain uncertain. The latest Plank results \citep{Plank2018} favor a reionization that happened late and fast ($z=7.82\pm0.71$), consistent with it being driven by photons from massive stars in low-mass galaxies \citep[e.g.][]{RobEll2015}, as long as the escape fraction of the ionizing radiation is high enough \citep[e.g.][]{Stark2016}.

In addition to star-forming galaxies, accretion disks of active galactic nuclei (AGN) are also possible sources of ionizing photons at high redshift \citep[e.g.][]{AroMcC1970,MeiMad1993}. Thus, they have long been considered possible contributors to reionization \citep[e.g.][]{GraGia2018}, or at least indirect factors in the reionization process \citep[e.g.][]{SeiHut2018,KakEll2018}. Such sources, however, are presently thought to play a minor role in the reionization of hydrogen \citep[e.g.][]{HopRic2007,OnoKas2017,ParDun2018,MatStr2018}.

Still, accretion might not be the only ionizing radiation source in AGN. In particular, the termination regions of AGN-produced jets are known to be filled with non-thermal electrons \citep{CroIne2018}, which cool efficiently through inverse Compton (IC) and synchrotron radiation. At such large distances from the jet base, and taking into account the high density of the cosmic microwave background (CMB) photon field at $z\simeq6$, it is expected that IC radiation will dominate radiative losses, upscattering CMB photons to higher energies \citep[e.g.][]{WuGhi2017}. 

For the brightest blazars there is evidence to indicate that jets may be as powerful as accretion radiation, if not more powerful \citep{GhiTav2014,SbaGhi2016}. \citet{SbaGhi2015} suggest that at $z\simeq6$ the jetted fraction of the most powerful AGN might be close to one. Whereas radiation from an accretion disk is easily absorbed by the dense obscuring medium surrounding the AGN, the jet lobes are located in regions free from dense surrounding material. Therefore, if the jetted fraction is high enough, the number of sources that contributed to reionization with photons from their jet lobes might be larger than those contributing with accretion disk photons. 

Recently, \citet{Bos2018} explored the possible role of AGN jets and their termination regions in the reionization epoch, using empirically derived black hole mass functions and assuming a certain duty-cycle and accretion power. The conclusion reached in that work was that jet lobes might contribute non-negligibly to the reionization of the Universe at $z\gtrsim 6$. 

In this work, we carry out a more quantitative study of the impact of AGN jet lobes in reionizing the Universe, expanding it up to significantly higher redshifts. To do that, we improve the estimations of \citet{Bos2018} using recent, empirically derived quasar luminosity functions (LFs) at $z\simeq6$, and correcting them for possible obscured sources. The work is structured as follows. In Sect.~\ref{LF} we discuss the adopted luminosity functions and how we correct them to account for the presence of obscured sources. In Sect.~\ref{Jets} we compute the fraction of jet power that goes to ionizing radiation. In Sect.~\ref{Reionization} we use the obtained results to estimate the contribution of AGN jets to the ionizing photon density at $z\simeq6$, and to the H II volume filling factor in the IGM in the range $z\simeq15-5$. Finally, we summarize and discuss our results in Sect.~\ref{Discussion}.

\section{Luminosity function}\label{LF}

In order to characterize the AGN population at the epoch of reionization we must assume a quasar LF. Various studies have attempted to construct the LF at $z\sim6$, including those based on optical-UV \citep[e.g.][]{WilDel2010,OnoKas2017,KulGir2018}, X-ray \citep[e.g.][]{ParDun2018,VitBra2018}, or radio data \citep[e.g.][]{CacMor2019}.

Luminosity functions derived from radio studies tend to be inconsistent with X-ray results, finding a lower density of sources as well as different density peaks as a function of redshift \citep[see, e.g.][]{AjeCost2009,CacMor2019}. \citet{WuGhi2017} and \citet{SaxRot2017} attribute the low number density of radio sources at $z>3$ to quenching of radio emission due to higher densities of the CMB (see also Sect.~\ref{Discussion}). On the other hand, X-ray studies are also inconsistent with those derived from rest-frame UV surveys, finding an excess of sources at lower luminosities. This is presumably associated with dust obscuration effects, which are much more important at UV wavelengths. However, the X-ray and radio LFs cover a broad redshift range, reaching much later times than we are interested in, and are generally derived using smaller source samples. We therefore opt to use the most recent UV results, and correct them for the effects of obscuration. In particular, we use the LF derived in one of the latest and most complete studies \citep{MatStr2018}, based on a compilation of rest-frame UV data \citep{JiaMcG2016,WilDel2010,MatStr2018}. Their sample has the advantage of covering a broad luminosity range, but within a narrow $5.7<z<6.5$ redshift range.

To the LF from \cite{MatStr2018} we added a correction for the absorbed AGN fraction based on results from \citet{VitBra2018}, who analyzed X-ray data of AGN in the $3<z<6$ range. They derive an obscured AGN fraction of $\approx 0.8$ at high X-ray luminosities, as well as a decrease in obscuration at $L_{\rm x}<10^{43}$~erg~s$^{-1}$. Although this decrease goes against the well-established trend that low-luminosity AGN are more frequently obscured than those of higher luminosity \citep[e.g.][]{Law1991,UedYos2003,SteBar2003,Sim2005}, \citet{VitBra2018} attribute this unexpected result to incompleteness of the sample at low luminosities, and thus determine it as unreliable. Based on this, we assumed first a constant obscuration (CO) fraction of $0.8$, a value that is compatible with their data in all luminosities in which the sample is complete. The LF we derive is
\begin{equation}\label{eq:LFmag}
\Phi_{\rm CO}(M_{1450})=\frac{\Phi_*}{10^{0.4*(\alpha_{\rm CO}+1)(M-M_*)}+10^{0.4*(\beta+1)(M-M_*)}}\,,
\end{equation}
in units of Gpc$^{-3}$~mag$^{-1}$, where $\alpha_{\rm CO}=-1.23$ and $\beta=-2.73$, $M_*=-24.9$ is the break magnitude, and $\Phi_*=5\times 10.9$ is the normalization corrected for an 80\% of obscured sources.

Nevertheless, as mentioned, there could be a trend in the obscured AGN fraction at $z\sim 6$ to increase at low luminosities, as confirmed by \citet{UedAki2014} for $z\lesssim 3$. This would imply that the less luminous sources in the sample could have obscured fractions larger than the assumed value of $0.8$, which could make our corrected LF conservative at low luminosities. We can account for this effect with a second correction, which we refer to as correction for differential obscuration (DO). Therefore, we consider that 80\% of the brightest observed sources ($L_{\rm bol}=10^{48}$ erg s$^{-1}$) are obscured and, following the trend derived by \citet{UedAki2014}, we then assume that sources three orders of magnitude fainter should be obscured about four times more often. This yields a second LF,$\Phi_{\rm DO}$, with the same parameters, but $\alpha_{\rm DO}=-1.76$.

\begin{figure}[t]
\centering
\includegraphics[width=0.49\textwidth,keepaspectratio]{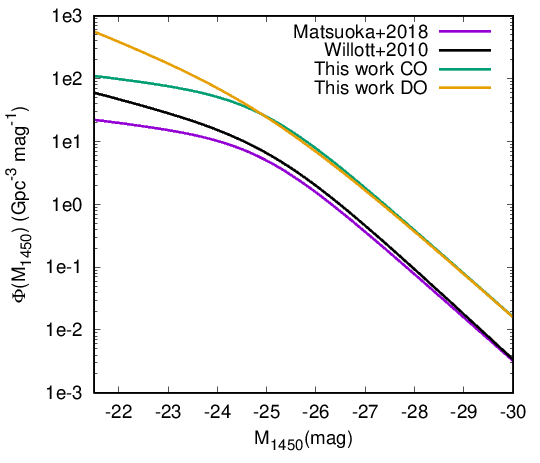}\hspace{20pt} 
\caption{Luminosity functions derived by \citet{MatStr2018} and \citet{WilAlb2010} compared to those used in this work. The two curves used here are based on that of \citet{MatStr2018} and include a constant correction for obscuration (CO) or a differential correction for obscuration (DO).}
\label{fig:LFmag}
\end{figure}

In Fig. \ref{fig:LFmag}, we compare our two LFs, CO and DO, with that originally derived by \citet{MatStr2018}, and that of \citet{WilAlb2010}. For further comparison, we also transformed the magnitude LFs, $\Phi(M_{1450})$, to bolometric luminosity using a correction factor of 4.4, as in \citet{WilAlb2010} \citep[from][]{RicLac2006}. We then transformed this to a black hole mass function (BHMF, $\Phi_{\rm BH}$). For this conversion, it is necessary to assume an Eddington ratio ($\lambda_{\rm Edd}\equiv L_{\rm bol}/L_{\rm Edd}$). For the sake of consistency, we used observational data taken from Fig.~3 of \citet{MatOno2019}, who analyzed the same sample of AGN used to derive our LFs. Using those data, we found an average $\lambda_{\rm Edd}=0.83\pm0.12$, and no significant trend with AGN luminosity or black hole mass. Averages in all the different luminosity bins taken are compatible, within the errors, with the average value for the whole sample.

\begin{figure}[t]
\centering
\includegraphics[width=0.5\textwidth,keepaspectratio]{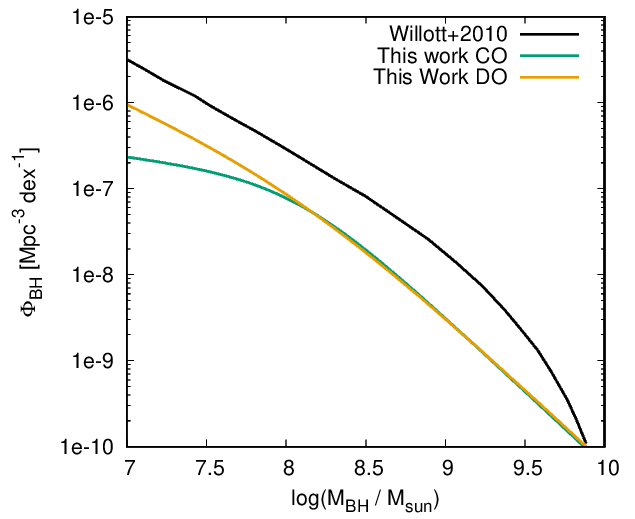}\hspace{20pt} 
\caption{Black hole mass functions derived from the luminosity functions in Fig. \ref{fig:LFmag}, compared to that of \citet{WilAlb2010}.}
\label{fig:BHMF}
\end{figure}

In Fig. \ref{fig:BHMF}, we compare the BHMFs derived in this work to that of \citet{WilAlb2010}, used by \citet{Bos2018} to obtain a first estimation of the contribution of AGN jets to reionization. Our BHMF values are lower than those of \citet{WilAlb2010}, partly because we do not account for the presence of quiescent black holes. In this sense, the BHMF from \citet{WilAlb2010}, among others, serves as a kind of upper limit; that is, our BHMFs should not exceed those including quiescent (or weakly accreting) AGN, as they are AGN-BHMFs. \citet{WilAlb2010} also used UV data to derive an observational LF, but instead of transforming it into a BHMF, they assumed a Schechter BHMF (hence the noticeably different shape), transformed the BHMF into an LF, and re-fitted it to the data. Other differences arise from small changes in the duty cycle and the assumed value of $\lambda_{\rm Edd}$, and a different correction prescription for obscured sources.

\section{Ionizing efficiency of jet lobes}\label{Jets}

The termination regions of AGN jets are expected to inflate lobes on scales of $\sim 100$~kpc, with the lobe pressure potentially dominated by non-thermal electrons. Moreover, the energetics of shocked shells of IGM at $z\gtrsim 6$ may be dominated by thermal cooling through free-free continuum and line emission. These factors could lead to a significant fraction of the jet luminosity being transformed into photons that would ionize, excite, and heat the IGM either through direct or indirect (via secondary electrons) interactions \citep[see][and references therein]{Bos2018}. 

We estimated the ionizing power of IC interactions between the relativistic electrons in the lobes, and CMB photons, which are upscattered into H-ionizing photons. We adopted different broad electron energy distributions and assumed a minimum particle energy of $E_{\rm e,min}=1$~MeV. In this scenario the energy injected into accelerating electrons in the jet would be turned into ionizing luminosity with an efficiency of $\approx 30-40$\% for $p\lesssim 3$ in a cooled electron energy distribution $\propto E^{-p}$ \citep[adopting the energy ratio going to ionization from][]{shull1985}. For $p>3$, the efficiency quickly goes down (e.g., 1\% for $p=4$), unless higher values of $E_{\rm e,min}$ are assumed, even above the minimum electron energy required to produce H-ionizing photons ($E_{\rm e,min}\approx 4\times 10^{-5}$~erg, or a Lorentz factor $\gamma_{\rm e,min}\approx 50$, at $z\simeq 6$). If this is the case, then higher efficiencies (up to $\sim 40$\%) can be reached regardless of the value of $p$. We note that \cite{WuGhi2017} adopt $\gamma_{\rm e,min}\sim 100$ for their modeling of lobe radio emission in high-$z$ blazars \citep[see sect.~3.2 in][]{Bos2018}.

\cite{Bos2018} concluded that the shocked IGM shell may be close to radiative. For a jet lobe suffering strong IC losses, the evolution of the shocked IGM shell formed by a jet with power $10^{44}$~erg~s$^{-1}$ (equivalent to a black hole mass $\simeq 10^6$~M$_\odot$ under our assumptions) would be likely radiative under primordial abundances. A larger, yet still relatively small, IGM metallicity, for example 1\% the solar value, would result in this shell evolution being even more radiative. For such a shocked IGM shell, the expected thermal-to-ionizing luminosity efficiency would be similar to that of IC, as the emission would be likely released in the far UV. However, a proper assessment of the thermal losses of the shocked IGM shell requires a detailed characterization of the jet lobe-IGM interaction (including IC losses), and some knowledge of the medium metallicity. 

\section{Contribution to reionization}\label{Reionization}

In this section we estimate the maximum contribution of AGN to reionization, both through their jets and through accretion onto the supermassive black hole. We assume that obscured sources do not contribute any UV photons to the IGM (unless jetted), and that unobscured sources are completely uncovered.

\subsection{Contribution of AGN jets to reionization}

The luminosity functions $\Phi_{CO,DO}(M_{1450})$ can be converted first into functions of luminosity, $L_{1450}$, and then into functions of bolometric luminosity, $L_{\rm bol}$, using the mentioned 4.4 correction factor. We assume that all AGN are jetted, extrapolating the results found by \citet{SbaGhi2015} for the few powerful blazars detected at high redshift at gamma-ray energies. Deviations from this assumption, and from the obscured fraction taken in Sect. \ref{LF}, are included within a parameter $\epsilon$. All numerical results presented in this section use a value $\epsilon=1$, which corresponds to a best-case scenario and should therefore be interpreted as an upper limit (see a discussion on this assumption in Sect. \ref{Discussion}).

It is then necessary to estimate how much energy goes into ionizing radiation as a function of $L_{\rm bol}$. First of all, we must assume a relation between accretion disk luminosity and jet power, $L_{\rm j}=\chi L_{\rm bol}$. There is evidence of a correlation between the two \citep[e.g.][]{RawSau1991,CelPad1997,GhiTav2010}, and \citet{GhiTav2014} find that the power of bright relativistic jets tends to be even higher than the luminosity of their accretion disks. We assume a value of $\chi=1$, and again all results on ionizing photon density scale with it. 

Following the results from Sect. \ref{Jets}, we assume that a factor $\xi=0.3$ of the jet power goes into ionizing radiation. This is the case if the lobe pressure is dominated by relativistic electrons that can produce H-ionizing photons via IC, and/or if the shocked IGM shell is radiative. The H-ionizing luminosity (comoving) density is then computed as
\begin{equation}\label{eq:epsilon}
\dot{\epsilon}_{\rm CO,DO}=\int \chi \ \xi \ \epsilon \ \Phi_{\rm CO,DO}(L_{\rm bol}) \ L_{\rm bol} \ {\rm d}L_{\rm bol}\,.
\end{equation}

We integrate in the range $L_{\rm bol}=10^{43}-10^{48}$~erg~s$^{-1}$, which corresponds to AGN with black hole masses in the range $M_{\rm BH}\approx 10^5-10^{10}$~M$_{\odot}$. The resulting values are $\dot{\epsilon}_{\rm CO}=4.3\times 10^{38}$~erg~s$^{-1}$~Mpc$^{-3}$ and $\dot{\epsilon}_{\rm DO}=9.2\times 10^{38}$~erg~s$^{-1}$~Mpc$^{-3}$. Considering  13.6~eV per H-ionizing photon, this translates to photon densities of $\dot{n}_{\rm CO/DO}=3.2\times10^{49},6.8\times10^{49}$~s$^{-1}$~Mpc$^{-3}$ (in the best-case scenario, $\chi=\xi=\epsilon=1$). 

\subsection{Contribution of AGN disks to reionization}\label{disk_corr}

\citet{MatStr2018} estimate the contribution of AGN accretion disks to reionization without correcting their $\Phi$ for the presence of obscured sources, as they assume no ionizing radiation can escape them. However, in a jetted source, a small fraction of ionizing radiation can escape in the direction of the jets (which must be unobscured) and contribute to the ionization of the surrounding medium. This would increase the contribution of disks in a factor $(1-f_{\rm obsc})^{-1}\epsilon f_{\rm esc}$, where $f_{\rm esc}$ is the escape fraction caused by the drilling of the jets and $f_{\rm obsc}$ the fraction of obscured sources (set to 0.8 in this work).  Considering $f_{\rm esc}=1$ for unobscured sources and $f_{\rm esc}=0.1$ for obscured ones, and using $\epsilon=1$ as is done for jets, AGN disks may contribute a 50\% more to $\dot{n}_{\rm ion}$ than accounted for by \citet{MatStr2018}. Using this corrected LF, the resulting ionizing photon density produced by the accretion disks of AGN is $\dot{n}_{\rm disk}=6.3\times10^{48}$~s$^{-1}$~Mpc$^{-3}$.

Other works, however, advocate for a much higher escape fraction of AGN disk photons. Strong AGN winds can penetrate through dense medium surrounding AGN nuclei, which may allow a fraction of the ionizing radiation to escape \cite[e.g.][]{Wagner2013,Menci2019} even in obscured sources. \cite{GraGia2018} analyzed a sample of 16 AGN at $z\sim$4, including both obscured and unobscured nuclei, and estimated an average escape fraction of ionizing radiation of $f_{\rm esc}=0.74$. 

It is unclear if extremely obscured (even Compton-thick) AGN at $z\sim$6 can have such large $f_{\rm esc}$ on average, due to the difficulty of accounting for high X-ray obscuration in a large number of sources without invoking large covering fractions. Recent observations of a high-$z$ source \cite[][at $z$=6.515]{Vito2019} can be interpreted as a highly X-ray obscured AGN that strongly emits in the UV, but the possibility of the absorbed X-rays originating in a much fainter companion cannot be ruled out.

However, given the observations of \cite{GraGia2018} at $z\sim$4 and the lack of complete, large X-ray samples at higher redshift, the possibility of large escape fractions for AGN disk photons cannot be completely ruled out either. We opt to include estimations using $f_{\rm esc}=0.74$ as an average value for all sources (obscured and unobscured) for both $\Phi_{\rm CO}$ and $\Phi_{\rm DO}$. In this way, the plots in Fig. \ref{fig:QHIIndot} show a range of possible values of the AGN disk contribution to reionization, from the small correction to the \cite{MatStr2018} LF to the large $f_{\rm esc}$ of \cite{GraGia2018}.

\subsection{Reionization at higher redshifts}

\begin{figure}[t]
\centering
\includegraphics[width=0.5\textwidth,keepaspectratio]{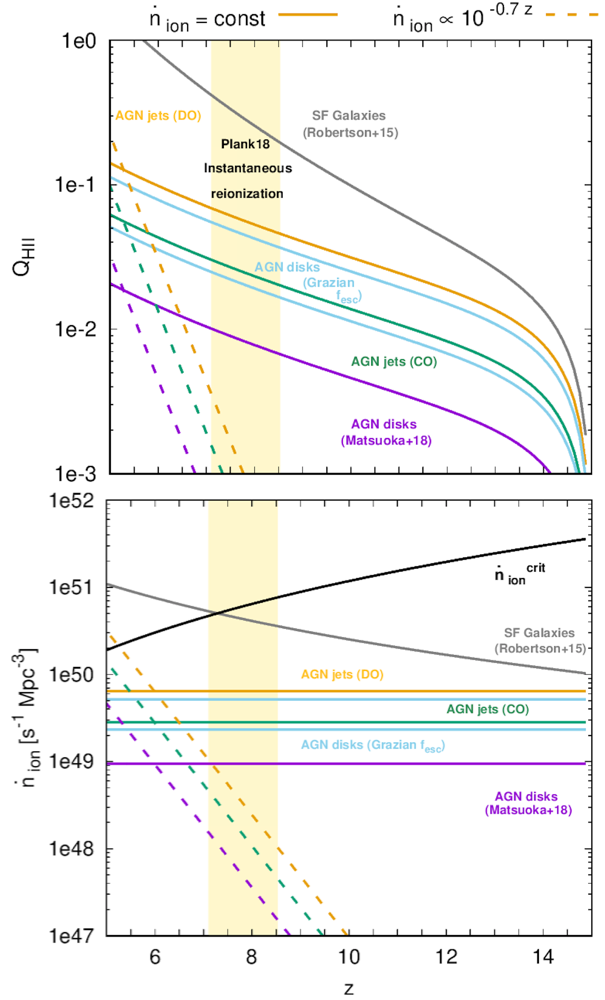}\hspace{20pt} 
\caption{Evolution of the H II volume filling factor (top) and the ionizing photon density (bottom) as a function of redshift. The plotted contributions are those of AGN disks \citep[][corrected for absorption]{MatStr2018}, star-forming galaxies \citep{RobEll2015}, and AGN jets (constant obscuration, CO, and differential obscuration, DO) for $\epsilon=1$ (maximum possible contributions). The blue lines are the contributions of AGN disks assuming an average $f_{\rm esc}=0.74$ as in \citet[][]{GraGia2018} for the CO (bottom blue line) and the DO (top blue line) normalizations. The shaded area represents the estimation of the redshift of instantaneous reionization \citep[$1\sigma$ confidence interval,][]{Plank2018}. The solid black line (bottom) represents the critical photon density necessary to keep the IGM ionized.}

\label{fig:QHIIndot}
\end{figure}

We can extrapolate our results to higher redshifts. The evolution of the H II volume filling factor in the IGM, $Q_{\rm HII}(t)$, is given by
\begin{equation}\label{eq:QHII}
\frac{{\rm d}Q_{\rm HII}}{{\rm d}t}=\frac{\dot{n}_{\rm ion}}{\overline{n}_{\rm H}} - \frac{Q_{\rm HII}}{\overline{t}_{\rm rec}}\,,
\end{equation}
where $\overline{n}_{\rm H}$ and $\overline{t}_{\rm rec}$ are the mean hydrogen density and recombination time, respectively \citep[see][]{MadHaa1999}. To numerically integrate this equation we consider that the IGM is fully neutral at $z\simeq15$, when reionization might have started \citep{BroYos2011,Dun2013}.

The ionizing photon densities necessary to balance recombination (in the ionized IGM, i.e., $Q_{\rm HII}=1.0$) is
\begin{equation}
\dot{n}_{\rm ion}^{\rm crit}= 10^{50.0} C_{\rm HII} \left( \frac{1+z}{7} \right)^3 {\rm s^{-1} Mpc^{-3}}\,,
\end{equation}
where $C_{\rm HII}$ is an effective HII clumping factor \citep{BolHae2007} within the range $C_{\rm HII}=1.0-5.0$ \citep{ShuHar2012}. We plotted a comparison between $\dot{n}_{\rm ion}^{\rm crit}$ and our derived $\dot{n}_{\rm ion}$ for two cases: assuming $\dot{n}_{\rm ion}$ constant with redshift, and assuming it evolves as $\propto 10^{-0.7 z}$ \citep[normalizing using the LF at $z\sim 6$;][]{MatStr2018} in Fig. \ref{fig:QHIIndot}. The figure also includes the evolution of the ionizing photon density generated by star-forming galaxies, $\dot{n}_{\rm stars}$ \citep{RobEll2015}, and that generated by AGN disks \citep[][corrected for absorption as in Sect. \ref{disk_corr}]{MatStr2018}. Figure \ref{fig:QHIIndot} also depicts the evolution of the HII volume filling factor as a function of redshift, including contributions from all the different considered $\dot{n}_{\rm ion}$. The clumping factor used in both plots is $C_{\rm HII}=3.0$, and varying it in the $1.0-5.0$ range can lead to relatively small variations \citep[see][]{MatStr2018}.

The plot for $\dot{n}_{\rm ion}$ shows that the contribution from star-forming galaxies is enough to fully maintain the IGM ionized at $z=6$. The contribution from AGN disks is likely small  (for low $f_{\rm esc}$), at most $\sim 6$\%, while AGN jets could contribute $\sim 10$\% (CO) or $\sim 20$\% (DO) in the best-case scenario. However, for $f_{\rm esc}=0.74$, AGN disks would contribute practically the same amount as their jets.

The plot for the $Q_{\rm HII}$ evolution shows, assuming a constant $\dot{n}_{\rm ion}$, that the contribution of AGN (through their jets, and/or their disks in the case of large escape fractions) to reionization could be $\sim 5$~\% (CO) or $\sim 10$~\% (DO) at $z=6$, the redshift at which the contribution of star-forming galaxies may suffice to fully reionize the IGM. When adopting $\dot{n}_{\rm ion}(z)\propto 10^{-0.7 z}$, derived from the $z$-evolution of the LF normalization at $z\lesssim 6$, we find that the contribution of AGN to $Q_{\rm HII}$ at $z\sim 6$ would be lower by a factor of $\sim3$.

\section{Summary and discussion}\label{Discussion}

We used LFs derived from UV data at $z\sim6$ to estimate the contribution of AGN jets to the reionization of the Universe. In order to do this, we took the LF of \citet{MatStr2018} and corrected it in three different ways to account for the presence of obscured sources. We considered an obscuration factor constant at all AGN luminosities (CO), fainter sources being more obscured than brighter ones (DO), and an LF with no turnover (NT). We estimated that $\sim30$\% of the jet energy of these sources could have turned into ionizing radiation. We computed the contribution of AGN jet lobes to the ionizing photon density and H~II volume filling factor at $z>6$.

\subsection{Contribution to reionization}

The contribution of star-forming galaxies, as derived by \citet{RobEll2015}, is both dominant and sufficient to reionize the Universe at $z\simeq6$, although their result depends on key assumptions that are still unconfirmed (see Sect. \ref{starforming}).

The contribution of jets to reionization is difficult to estimate without a precise knowledge of the evolution of the LF normalization at high redshifts (i.e., $\dot{n}_{\rm ion}(z)$). It might be negligible if we assume a strong decay with redshift, or it might be as high as $\gtrsim 10$\% if it remains constant, the jetted AGN fraction is close to 1, and the fraction of obscured sources is high.

Our results indicate that AGN jet lobes could generate as much as $\sim20$\% of the necessary photons to keep the IGM ionized at $z=6$ \citep[$\dot{n}_{obs}\sim 7 \times 10^{50}$ s$^{-1}$ Mpc$^{-1}$,][]{Mad2017}, well above the minimum $\sim6$\% derived for accretion disks, due to the lack of obscuration effects in the jet lobe scenario. We note, however, that $f_{\rm esc}$ is not a fully constrained parameter (see Sect. \ref{disk_corr}). Depending on its value, the contribution of AGN disks may be as high as the upper limits we estimate for the jet radiation. In this scenario their combined contributions would be non-negligible compared with that of star-forming galaxies.

However, the high reionization impact of jets requires assuming a number of things: first, either a large relativistic electron pressure in the lobes or radiative shocked IGM shells, or both; second, a jetted AGN fraction of almost 1; and third, that our estimation of the number of obscured sources at $z\sim6$ is correct. Again, if these conditions are not met, their contribution becomes negligible (see Sect. \ref{errors} for a discussion).

In addition to these considerations, different approaches to estimate the ionization power of AGN jets can render relatively different results. For instance, using the BHMF derived by \citet{WilAlb2010}, \citet{Bos2018} derived $\dot{n}_{\rm ion}=1.5\times10^{50}$~s$^{-1}$~Mpc$^{-3}$ in a best-case scenario, a factor of $\sim 2$ larger than our $\dot{n}_{\rm DO}$. Similarly, considering a higher minimum AGN luminosity (i.e., $L_{\rm j}=10^{45}$~erg~s$^{-1}$, equivalent to $M_{\rm BH}\simeq 10^7$ M$_{\odot}$, which corresponds to the faintest data points in the sample used to derive the LFs) can lead to significant changes. The contribution of both AGN disks and of $\Phi_{\rm CO}$ to reionization would vary only by $\sim10$\%, and that of $\Phi_{\rm DO}$ by a factor of $\sim2$. However, it is unlikely that no AGN of lower luminosities exist, and one must take into account that the contribution of those sources may be significant. Remarkably, changes in $\alpha$ (in this work, $\alpha=-1.23,-1.76$), the least constrained of the LF parameters, yield very different results.

\subsection{Jetted fraction and obscured sources}\label{errors}

In addition to $\alpha$, another important and not fully constrained parameter is the normalization (i.e., total number of sources). The results presented in Sect. \ref{Reionization} all scale linearly with $\epsilon$. Our assumption of $\epsilon=1$ implies that all AGN at $z\simeq6$ are jetted. \citet{SbaGhi2015} analyzed \textit{Swift} data of known $z>4$ blazars (i.e., five sources with $L_{\rm j}\sim10^{47}$ erg s$^{-1}$) and suggested that jetted sources might be enough to represent all AGN at those redshifts. This conclusion strongly depends on the derived Lorentz factors of the blazars, and the small statistics imply a large uncertainty on the jetted source fraction. Also, whether their results hold for lower luminosity AGN is uncertain.

This $\epsilon$ should also include possible deviations from the assumed 80\% of obscured sources. Whereas \citet{VitBra2018} find this value in the $z=3-6$ range, most of their sources have $z<4$. The maximum obscuration fraction may be expected at $z\sim2-3$, at the peak of star formation in the Universe. Therefore, while \citet{VitBra2018} do not observe this trend, a decay of obscuration fraction at $z\simeq6$ is possible. 

Our results, therefore, all scale with 
\begin{equation}
\epsilon=f_{\rm jet} \left( \frac{1-f_{\rm obsc}}{1-0.8}\ \right)^{-1},
\end{equation}
where $f_{\rm jet}$ and $f_{\rm obsc}$ are the jetted and obscured AGN fractions, respectively. If $\epsilon<0.2$, AGN jets would be contributing to reionization less than accretion disks at their lowest possible contribution; for example, with $f_{\rm obsc}=f_{\rm jet}=0.5$, $\epsilon=0.2$ is already reached. 

We note that studies of X-ray binaries show that their jets are produced under certain conditions of accretion (advection dominated), at either very low rates $(\lambda_{\rm Edd}\sim 0.1$) or very high rates ($\lambda_{\rm Edd}\sim1$), and that otherwise disk emission dominates \citep[e.g.][]{Fender2004}. The Eddington ratios of $z\sim6$ quasars of \cite{MatOno2019} are distributed around the mean value of $\sim0.8$. If the behavior of SMBH at high $z$ depended on accretion in a similar manner, the jetted AGN fraction could be low. However, it is unclear whether the behavior of X-ray binaries can be extrapolated to SMBH at high $z$. 

We note that the X-ray data of \citet{VitBra2018} only extends up to AGN with $L_{\rm bol}=10^{47}$ erg s$^{-1}$, and, although they do not observe a clear trend with luminosity, assuming $L_{\rm bol}=10^{48}$ erg s$^{-1}$ to have an obscured fraction of $0.8$ may be an overestimation. However, redoing the calculations excluding the most luminous AGN ($L_{\rm bol}=10^{47}-10^{48}$ erg s$^{-1}$) results only in a decrease in $Q_{\rm HII}$ of $\sim10$\% when using $\Phi_{\rm CO}$ and a negligible decrease when using $\Phi_{\rm DO}$.

On another note, the limited sensitivity of the current surveys means that we have no accurate knowledge of the number of low-luminosity AGN at high $z$. For instance, intermediate-mass black holes in the center of gas-rich dwarf galaxies may be active at $z\gtrsim 6$ as mechanical feedback could shutter both star formation and AGN activity \citep[e.g.][]{silk17}. Also, weakly accreting black holes of any mass could contribute to reionization to some extent, but would be unnoticed by observations. Finally, the present observational constraints on black hole past activity (e.g., accreted mass, accretion rate) do not allow the derivation of strong constraints on the ionizing contribution of AGN jets at very high $z$ \citep[see][and references therein]{Bos2018}.

\subsection{Possible incompletness in AGN color selection}\label{normalization}

In a recent study of AGN selection at $z\sim4$, \cite{Boutsia2018} find that out of their 16 spectroscopically confirmed AGN only 6 were selected by color. They argue, based on their results, that selections using solely color criteria can be highly incomplete (at a level of $\sim$50\%), particularly for faint sources. The selection used by \cite{MatStr2018}, which we use to derive our LF, is also based on color and therefore could be affected by a similar incompleteness effect.

Given how the color selection used by \citet{Boutsia2018} and \citet{MatStr2018} is in different bands (due to the distinction between $z\sim 4$ and $z\sim 6$ selection) and, more importantly, the possible overlap between the incompleteness correction and the correction for obscured sources ($f_{\rm obsc}$), we have opted to not apply this factor of 2 increase in the LFs used in this work. However, we note that this effect could be present in the \cite{MatStr2018} sample, and that if that were the case all the results presented here could be increased by a factor of up to $\sim$2 (both for the AGN jet and for disk contributions to reionization).

\subsection{Contribution of star-forming galaxies}\label{starforming}

The star formation contribution to the high-$z$ UV background and reionization depends mainly on the total star formation rate density and on the escape fraction of UV photons from the star formation sites. The curves plotted in Fig. \ref{fig:QHIIndot} of this work are derived by \citep{RobEll2015}, who assume an escape fraction of ionizing radiation $f_{\rm esc}\simeq0.2$ and extrapolate the LF below the observed limits. 

Using their model, the minimum galaxy luminosity required to achieve reionization within the Plank-derived redsfhit limits is $M_{\rm UV} \simeq -13$ \citep[much fainter than the current detection limits,][]{Stark2016}. There is also the possibility of an  accelerated decline in $\rho_{SFR}(z>8)$ \citep[e.g.][]{OesBou2014}, which would reduce their impact on reionization. Ultradeep infrared imaging with JWST is necessary to provide robust constraints on the shape of the UV luminosity function at luminosities below $M_{\rm UV} \simeq -17$.

An escape fraction of $f_{\rm esc}\simeq0.2$ is also necessary to achieve the ionizing photon densities required to reionize the IGM. There are indications that the escape fraction is larger in low-luminosity galaxies at $z>3$ \citep{Nestor2013} and may increase with redshift at $z>3$ \citep{Jones2013}, which suggests that such large escape fractions may not be unreasonable. However, other studies indicate that it is not easy for the high-$z$ star-forming galaxy population to reach $f_{\rm esc}\simeq0.1$ \citep[][and references therein]{Grazian2017}. Also, a recent work on
faint galaxies hosting high-$z$ Gamma Ray Bursts finds extremely small escape fractions \citep{Tanvir2019}, adding to the above difficulty.

Accurately quantifying the contribution of star-forming
galaxies to the reionization of the Universe thus still appears rather difficult today.

\subsection{Quenched radio emission}

As mentioned in Sect.~\ref{LF}, there is a discrepancy in the LF at different energy bands, with radio LFs, which should account for jetted sources, finding lower densities of AGN at high redshift than those derived at high energies. It is natural, however, to expect significantly less synchrotron emission with respect to IC emission in the extended jet regions as the CMB energy density is $\propto (1+z)^4$. The reason is that unless radiation comes from very close to the jet base, synchrotron emission is suppressed at high redshift. This can take place in two different contexts. First, non-radiative losses can be dominant (e.g., adiabatic losses due to the jet expansion). Since the photon energy density of the CMB is higher than the energy density of the magnetic field, IC emission can be much brighter than synchrotron emission. To exemplify this, we can consider the particular case of a jet with a total power of $10^{44}$~erg~s$^{-1}$, a Lorentz factor of 10, Poynting flux equal to 10\% of the matter energy flux, and a half-opening angle of 0.1~rad, at $z\sim6$. In such a case, the synchrotron emission can only overcome the CMB IC luminosity at a distance $\lesssim10$~pc from the jet base. The same jet in the local universe could have a synchrotron component brighter than the IC component up to a jet height of $\sim 300$~pc. This effect leads to comparatively stronger IC emission. Second, IC emission might be so intense that it would dominate over non-radiative losses, with radio electrons losing most of their energy via IC CMB. This effect reduces  the radio emission with respect to the case with dominant non-radiative losses. Accounting for these effects, one may easily expect different LF $z$-evolutions at different frequencies.

\citet{WuGhi2017} study this mechanism for radio quenching at $z>3$ and conclude that it can efficiently dim the diffuse radio emission from jetted AGN. However, their limited sample does not allow them to confirm whether the mechanism is entirely sufficient to explain the radio-loud AGN deficit at high redshifts.  

Finally, it is worth mentioning that, in addition to ionizing the medium, about 10--20\% of the energy of jet lobe hard photons may go to heating the IGM at $z\gtrsim 6$. An accurate estimate of the level of IGM heating due to jet lobes is beyond the scope of this work, but certainly it should be compatible with the thermal history of the IGM at very high redshift \citep[see, e.g.,][]{daloisio2017,garaldi2019}.

\section*{Acknowledgments}
We thank the anonymous referee for a careful reading of the manuscript and for useful comments and suggestions that improved the paper. We acknowledge support by the Spanish Ministerio de Econom\'{i}a y Competitividad (MINECO/FEDER, UE) under grant AYA2016-76012-C3-1-P, with partial support by the European Regional Development Fund (ERDF/FEDER), MDM-2014-0369 of ICCUB (Unidad de Excelencia `Mar\'{i}a de Maeztu'), and the Catalan DEC grant 2017 SGR 643. N.T-A. acknowledges support from MINECO through FPU14/04887 grant. 
\bibliographystyle{aa}
\bibliography{Referencies}   


\end{document}